\def\imat  {i} 
\documentstyle[twocolumn,floats,epsf,ifthen,aps,prb]{revtex} 

\begin{document} 

 

\title{Simple quantum feedback of a solid-state qubit} 

\author{Alexander N. Korotkov} 
\address{ Department of Electrical Engineering, 
University of California, Riverside, CA 92521-0204 
} 
\date{\today} 
 
\maketitle 
 
\begin{abstract} 
   We propose an experiment on quantum feedback control of 
a solid-state qubit, which is almost within the reach of the 
present-day technology. Similar to the earlier proposal, 
the feedback loop is used to maintain the coherent (Rabi) 
oscillations in a qubit for an arbitrary long time; however, this is done 
in a significantly simpler way, which requires much smaller 
bandwidth of the control circuitry. 
The main idea is to use the quadrature components 
of the noisy detector current to monitor approximately the phase
of qubit oscillations. 
  The price for simplicity is a less-than-ideal operation: 
the fidelity is limited by about 95\%. The feedback loop operation 
can be experimentally verified by appearance of a positive in-phase 
component of  
the detector current relative to an external oscillating signal
used for synchronization. 
\end{abstract} 
 
 

The needs of quantum computing \cite{Nielsen} are fueling a rapid 
progress in experiments with solid-state qubits.
In particular, quantum coherent (Rabi) oscillations have been 
demonstrated using superconducting charge, 
flux, and phase qubits \cite{sc-qubits} as well as double-quantum-dot
qubits.\cite{dqd-qubit} 
Successful experiments with two superconducting qubits have also been 
demonstrated. \cite{two-qubit} Even though at present only
very basic operations with qubits are experimentally accessible, 
more advanced experiments are a natural next stage. 
One of the directions for the advanced qubit control is 
realization of the quantum feedback control of a solid-state 
qubit, \cite{Ruskov-fb} which can be used in a quantum computer
for qubit initialization and is also an important demonstration
by itself, clarifying the controversial issue of gradual collapse 
of a quantum state. (In optics quantum feedback control was proposed
more than a decade ago \cite{Wiseman-93-fb} and has been already 
demonstrated experimentally.\cite{Armen}) 

    For the analysis of a quantum feedback we have to take into account
the process of continuous qubit collapse. Therefore, the conventional
approach to continuous quantum measurement \cite{Caldeira,Makhlin} 
is inapplicable, and it is necessary to use the recently developed Bayesian 
approach \cite{Kor-Bayes} or the equivalent (though much different 
technically) approach of quantum trajectories.\cite{Goan} 
The possibility of a quantum feedback is based on the fact that 
measurement by an ideal solid-state detector 
(with 100\% quantum efficiency $\eta$) 
 does not decohere a single qubit, \cite{Kor-Bayes} 
even though it decoheres an ensemble of qubits 
because each qubit evolves in a different way. 
The random evolution of a qubit in the process of measurement 
can be monitored using the noisy detector output, with accuracy 
depending on $\eta$, so that for an ideal detector ($\eta=1$) 
even the monitoring of qubit wavefunction is possible. An example 
of theoretically ideal solid state detector is \cite{Aleiner,Kor-Bayes} 
the  quantum  point  contact  ($\eta$  comparable  to  1  has  been 
demonstrated
experimentally \cite{Buks}). The single-electron transistor is 
significantly nonideal \cite{Makhlin,Kor-Bayes,Devoret} ($\eta \ll 1$) 
in the semiclassical ``orthodox'' mode of operation; \cite{Av-Likh}
however, it can reach ideality in some modes based on cotunneling or
Cooper pair tunneling. \cite{SET-ideal}

        Monitoring of the quantum state in real time can naturally be 
used for continuous feedback control of a quantum system. 
        In the experiment proposed in Ref.\ \cite{Ruskov-fb} the quantum 
feedback is used to maintain quantum coherent (Rabi\cite{Rabi}) 
oscillations in a qubit for an arbitrary long time, synchronizing them
with an external classical signal. 
This is done by measuring the noisy current $I(t)$ in a weakly 
coupled detector and using the quantum Bayesian equations \cite{Kor-Bayes} 
to translate information contained in $I(t)$ into the evolution of 
qubit density matrix $\rho (t)$. After that $\rho (t)$ is compared with 
the desired quantum state $\rho_d(t)$, and the calculated difference is used 
to control the qubit Hamiltonian in order to decrease the difference. 
Notice that the measurement backaction necessarily shifts the phase 
of Rabi oscillations in a random way (adding it to dephasing due to 
environment); however, the information contained 
in $I(t)$ is sufficient to monitor this change and therefore restore 
the desired phase.

    An important difficulty in such experiment is a necessity to solve
the Bayesian equations in real time. Moreover,
the bandwidth of the line delivering $I(t)$ to the circuit solving
the Bayesian equations, should be significantly wider than the Rabi
frequency $\Omega$ (otherwise the information contained in the noise
is lost). Unfortunately, these conditions are unrealistic for the 
present-day experiments with solid-state qubits. (The ``direct'' feedback
also analyzed in Ref.\ \cite{Ruskov-fb} does not require solving
Bayesian equations, but requires a wide bandwidth of the whole feedback
loop.) 

\begin{figure} 
\centerline{
\epsfxsize=2.9in 
\vspace{0.1cm}
\epsfbox{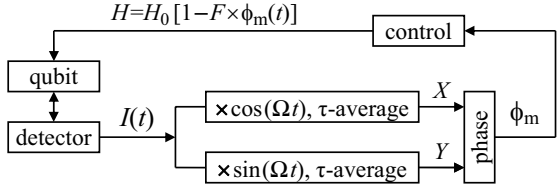} 
} 
\vspace{0.2cm} 
\caption{Schematic of the proposed quantum feedback loop. Two quadrature 
components of the detector current $I(t)$ are used to monitor 
approximately the phase difference between Rabi oscillations and a local 
oscillator, which is used to control the qubit parameter $H$. The phase 
can also be monitored using a tank circuit. Positive average in-phase
quadrature $\langle X \rangle$ is an experimental indication of 
quantum feedback operation. 
 }
\label{schematic}\end{figure}

    In this paper we propose and analyze a much simpler way 
(Fig.\ \ref{schematic}) 
of processing the information carried by the detector current $I(t)$. 
The idea is to use the fact that besides noise, $I(t)$ contains 
an oscillating contribution due to Rabi oscillations in the measured qubit.
  Therefore, if we apply $I(t)$ to a simple tank circuit 
(which is in resonance with $\Omega$), then the phase of the tank circuit 
oscillations will depend on the phase of Rabi oscillations. 
Instead of using the tank circuit, almost equivalent theoretically 
procedure is to mix $I(t)$ with the signal from a local oscillator 
(Fig.\ \ref{schematic}) in order to determine two quadrature 
amplitudes of $I(t)$ at frequency $\Omega$, which will carry information 
on the phase of Rabi oscillations. 
  Since diffusion of the Rabi phase is a slow process (assuming weak
coupling to the detector and environment), the further circuitry 
can be relatively slow, limited by the qubit dephasing rate, 
but not limited by much higher Rabi frequency. 
The simplicity of the information 
processing and relatively small required bandwidth are the main advantages
of this proposal in comparison with Ref.\ \cite{Ruskov-fb} 
    The experiment can be realized using either superconducting 
\cite{sc-qubits,two-qubit,LaHaye} or GaAs \cite{dqd-qubit,Buks} technology.

        The idea of this proposal partially stems from the fact 
that in absence of feedback the qubit Rabi oscillations lead to a noticeable 
peak in the spectral density $S_I(\omega )$ of the detector current 
at $\omega \approx \Omega$, with the peak-to-pedestal ratio up to
4 times \cite{factor4} (somewhat similar 
experiments have been reported recently \cite{Il'ichev}).
Since 4 is not a big number, one would expect quite inaccurate phase 
information carried by current quadratures and therefore poor operation of
the feedback. Surprisingly, the quantum feedback operates much better 
than it would be expected from classical analysis.

        Let us consider a ``charge'' qubit (either double quantum dot or 
single Cooper pair box) with Hamiltonian 
$   {\cal H}_{qb} = (\varepsilon /2) (c_2^\dagger c_2 - c_1^\dagger c_1)
+H (c_1^\dagger c_2 + c_2^\dagger c_1)$, 
   where $c_{1,2}^\dagger$ and $c_{1,2}$ are the creation and annihilation
operators in the basis of ``localized'' (charge) states, 
$\varepsilon$ is their energy asymmetry, and the tunneling 
$H=H_0+H_{fb}(t)$ can be controlled by the feedback loop ($H_{fb}$). 
We assume the standard coupling \cite{Gurvitz,Kor-Bayes,factor4} 
between the 
charge qubit and the detector (quantum point contact or single-electron 
transistor). Instead of writing Hamiltonians explicitly, we will 
characterize the measurement by two levels of the average detector current, 
$I_1$ and $I_2$, corresponding to the two charge states, by the 
detector output noise $S_I$, and by the total ensemble-averaged qubit 
dephasing rate $\Gamma$ due to detector backaction and environment. 
Assuming sufficiently large
detector voltage and quasicontinuous detector current $I(t)$, we 
describe the qubit evolution by the Bayesian equations \cite{Kor-Bayes}
(in Stratonovich form) 
        \begin{eqnarray} 
&&  \dot{\rho}_{11}= 
-2 H\,\mbox{Im}\,\rho_{12}
         +2 \rho_{11}\rho_{22} [I(t)-I_0]\, \Delta I/S_I, 
        \label{Bayes1}\\ 
&& {\dot\rho}_{12}=  \imat \varepsilon \rho_{12}+ 
        \imat H  (\rho_{11}-\rho_{22}) -\gamma \rho_{12} 
       \nonumber \\ 
&& \hspace{0.8cm}  -( \rho_{11}-  \rho_{22})   \rho_{12} 
[I(t)-I_0]\,  \Delta I/S_I  \, ,      
        \label{Bayes2} 
        \end{eqnarray}
where $\hbar =1$, $\Delta I=I_1-I_2$, $I_0=(I_1+I_2)/2$, and 
$\gamma =\Gamma -(\Delta I)^2/4S_I$. The qubit decoherence rate 
$\gamma = \gamma_d +\gamma_e$ is due to detector nonideality,
$\gamma_d = (\eta^{-1}-1)(\Delta I)^2/S_I$, and due to additional coupling 
with environment ($\gamma_e$). The current $I(t)=I_0+(\rho_{11}-\rho_{22})
\Delta I/2 +\xi (t)$ has the noise component $\xi (t)$ with the flat (white) 
spectral density $S_I$.
Notice that in the case $\varepsilon =0$ (which is assumed unless mentioned 
otherwise), 
we can disregard the evolution of 
$\mbox{Re} \rho_{12}$ (it becomes zero at $t\gg \Gamma^{-1}$), so
only two degrees of freedom are left, which may be parameterized as 
$\rho_{11}-\rho_{22}=P \cos (\Omega t +\phi)$ and 
$2 \mbox{Im} \rho_{12} =P \sin (\Omega t +\phi)$, where the 
feedback-maintained frequency $\Omega$ (see below) is assumed to be
equal (unless stated otherwise) to the bare Rabi frequency
$\Omega_0 =(4H_0^2 +\varepsilon^2)^{1/2}=2H_0$.
Moreover, in the ideal case $\gamma =0$ the state eventually 
becomes pure,\cite{Kor-Bayes} so that $P=1$ and the evolution can be 
described by only one parameter $\phi (t)$.

        We assume that two quadrature components of the detector current 
(Fig.\ \ref{schematic}) are determined as
        \begin{eqnarray}
X(t)=\int^t_{-\infty} [I(t')-I_0] \cos (\Omega t') e^{-(t-t')/\tau} dt'\, ,
\label{Xdef}\\ 
Y(t)=\int^t_{-\infty} [I(t')-I_0] \sin (\Omega t') e^{-(t-t')/\tau} dt' \, ,
\label{Ydef}    \end{eqnarray}
where $\Omega$ is the local oscillator frequency applied to the mixer,
and $\tau$ is the averaging time constant. Similar formulas are also
applicable to the case of a tank circuit with the resonant frequency 
$\Omega$ and quality factor $Q=\Omega\tau /2$.
If the detector current would be a harmonic signal 
$I(t)=I_0+ P(\Delta I/2) \cos (\Omega t+\phi_0 )$, then 
$\phi_0 = -\arctan (\langle Y\rangle / \langle X\rangle )$, 
so it is natural to use
\begin{equation}
 \phi_m (t) \equiv -\arctan (Y/X)
\end{equation} 
as a monitored estimate of the phase shift $\phi (t)$ 
between the Rabi oscillations and the local oscillator ($\langle \dots
\rangle$ means averaging over time). 

\begin{figure} 
\centerline{
\epsfxsize=2.9in 
\vspace{0.1cm}
\epsfbox{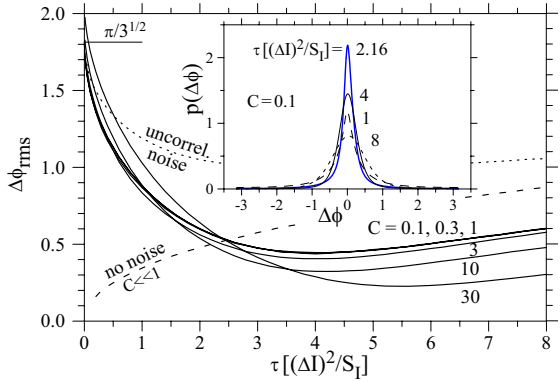} 
} 
\caption{Dependence of monitoring inaccuracy $\Delta \phi_{rms}$ on averaging
time $\tau$ without feedback for several values of coupling ${\cal C}$. 
Dashed and dotted lines correspond to classical signals. 
Inset: distribution of $\Delta \phi$ for several $\tau$ at weak coupling.
 }
\label{dphi-tau}\end{figure}

   Let us assume $\gamma =0$ and analyze first how close is the estimate 
$\phi_m (t)$ to the actual phase $\phi (t)$ without feedback, 
in which case $\phi$ evolves in a diffusive manner due to detector 
backaction. Figure \ref{dphi-tau} shows  
the rms phase difference $\Delta \phi_{rms}=\langle (\phi_m -\phi )^2 
\rangle ^{1/2}$ (solid lines) as a function of $\tau$ for several 
values of the 
dimensionless qubit-detector coupling ${\cal C}\equiv (\Delta I)^2/S_IH_0$, 
calculated numerically using Monte Carlo simulation of the measurement 
process.\cite{Kor-Bayes} At weak coupling, 
${\cal C} \lesssim 1$, the curves practically coincide, 
and the minimum $\Delta \phi_{rms} \approx 0.44$ is 
achieved at $\tau \approx 4S_I/(\Delta I)^2= 1/\Gamma$, 
which is expectable since $\Gamma$ determines the phase diffusion:
\cite{Kor-Bayes,Goan,factor4} 
$\langle [\phi (t)-\phi(0)]^2\rangle /t =\Gamma $. 
 At larger $\tau$, $\phi_m$ includes too much of 
irrelevant information from distant past, while at smaller $\tau$ the 
quadrature
amplitudes suffer too much from the noise. At $\tau \rightarrow 0$ (as well
as at $\tau \rightarrow \infty$) $\Delta \phi_{rms} \rightarrow 
\pi /\sqrt{3} \approx 1.81$ 
that corresponds to the uniform distribution of $\Delta \phi =\phi_m-\phi$ 
(complete absence of correlation between $\phi$ and $\phi_m$) within 
$\pm \pi$ interval (all phases are defined
modulo $2\pi$). 

        It it important to notice that the calculated $\Delta \phi_{rms}$ 
is significantly smaller than for a naive classical case, in which the 
noise $\xi (t)$ is not correlated with diffusive evolution of $\phi$. 
The dotted line in Fig.\ \ref{dphi-tau} shows the result for such a 
case at weak coupling, which has a minimum $\Delta \phi_{rms} \approx 1.0 $
[actually, for this curve we even increased the signal, assuming 
$I(t)-I_0=\sqrt{2}(\Delta I/2)\cos (\Omega t+\phi ) +\xi (t)$, 
which corresponds to correct spectrum\cite{factor4}]. 
Even more surprisingly, at $\tau > 2.5 S_I/(\Delta I)^2$ the 
inaccuracy $\Delta \phi_{rms}$ in the quantum case is smaller than for
the classical noiseless case, $\xi (t)=0$ (dashed line), which means
that the {\it noise improves the monitoring accuracy}. This quantum behavior
can be understood from the phase evolution equation
\cite{Ruskov-fb,Kor-Bayes}
        \begin{equation}
\dot \phi =- [I(t)-I_0] \sin (\Omega t+\phi) (\Delta I/S_I)+\Omega_0-\Omega .
        \label{dot-phi}\end{equation}
This equation \cite{witheta} shows that the quadrature component 
of the noise which shifts the observed phase $\phi_m$, also shifts the 
actual phase $\phi$ in the same direction. In other words, when the noise 
imitates oscillations, it forces the real Rabi oscillations to evolve 
closer to what is observed.

Inset in Fig.\ \ref{dphi-tau} shows the distribution of $\Delta \phi$ 
in the weak coupling limit for several values of $\tau$. 
The distributions are significantly non-Gaussian 
with the central part significantly 
narrower then $\Delta \phi_{rms}$. 
It is interesting that the value $\tau =4 S_I/(\Delta I)^2$ corresponding
to the minimum $\Delta \phi_{rms}$, does not provide the highest peak
of the $\Delta \phi$ distribution.
 To find the best $\tau$ in this respect, let us compare 
the monitored phase evolution 
$\dot\phi_m =- [I(t)-I_0]\sin (\Omega t+\phi_m)/(X^2+Y^2)^{1/2}$, 
with Eq.\ (\ref{dot-phi}). 
We would expect the best approximation of $\phi$ by $\phi_m$ 
when $\langle X^2+Y^2\rangle = (S_I/\Delta I)^2$. Using the definitions
(\ref{Xdef})--(\ref{Ydef}) and the current-current correlation function
\cite{factor4} $\langle I(0)I(t)\rangle =(S_I/2)\delta (t) +
(\Delta I/2)^2 \cos (\Omega t) \exp [-(\Delta I)^2t/8S_I]$, we obtain
$\langle X^2 +Y^2\rangle = S_I\tau [1/4+1/(1+8S_I/(\Delta I)^2\tau )]$ 
at $\Omega \tau \gg 1$ and ${\cal C}\ll 1$, 
so the condition $\langle X^2+Y^2\rangle =(S_I/\Delta I)^2$ 
is satisfied at $\tau (\Delta I)^2/S_I=(2/5)(\sqrt{41}-1) \approx 2.16$. 
This indeed corresponds to the largest peak of $\Delta \phi$ distribution
(see inset in Fig.\ \ref{dphi-tau}).

\begin{figure} 
\centerline{
\epsfxsize=2.7in 
\vspace{0.1cm}
\epsfbox{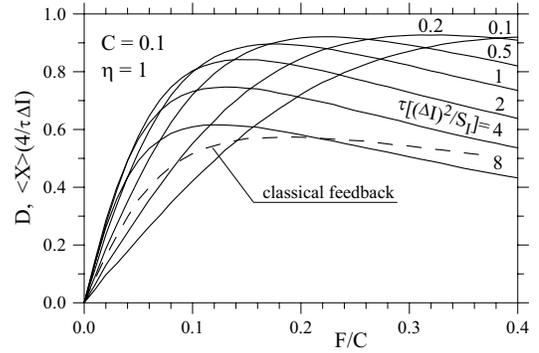} 
} 
\caption{Dependence of the synchronization degree $D$ on the feedback
factor $F$ in ideal case ($\gamma =0$) for several $\tau$. Experimentally
$D$ can be measured via average in-phase current quadrature 
$\langle X\rangle$. Dashed line is for a classical feedback.
 }
\label{D-F}\end{figure}

        Reasonably small difference between $\phi$ and $\phi_m$ in absence
of feedback implies that we can expect decent operation of the quantum 
feedback loop in which the phase estimate $\phi_m$ is used for
determining the feedback action. 
 Similar to Ref.\ \cite{Ruskov-fb} we 
consider the feedback loop, which aim is to suppress the fluctuations of
the Rabi phase, so that the goal is $\phi (t)=0$ (or as small as possible). 
It has been shown that this goal can be fully reached using 
the linear feedback rule $H_{fb}(t)/H_0 =-F \phi (t)$, which requires exact 
monitoring of $\phi$; here we analyze the operation of the feedback loop
with $H_{fb}/H_0= -F \phi_m (t)$, where $F$ is the dimensionless 
feedback factor (by definition $|\phi_m|<\pi$).

        We will characterize the performance of the feedback loop 
by the synchronization degree $D=\langle P(t)\cos \phi(t)\rangle =
2{\cal F} -1$ where ${\cal F}=\langle \mbox{Tr} \rho (t) \rho_d(t)\rangle$ 
is fidelity and 
 $\rho_d$ corresponds to the desired perfect Rabi oscillations 
($P_d=1$, $\phi_d=0$). 
        Figure \ref{D-F} shows (solid lines) the dependence of $D$ on 
the feedback factor $F$ for several time constants $\tau$ in the case of weak
coupling ${\cal C}=0.1$ and $\gamma =0$ (we normalize $F$ by ${\cal C}$, so 
the results practically do not depend on ${\cal C}$ for ${\cal C}\lesssim 1$).
\cite{H/Hfb} 
 One can see that each curve has a maximum, so that
the ``oversteering'' effect at larger $F$ makes the feedback performance
worse (this is in contrast to the case of Ref.\ \cite{Ruskov-fb} in which 
larger $F$ was always better). Somewhat unexpectedly, $\tau =1/\Gamma
=4S_I/(\Delta I)^2$ 
is no longer an optimum, and smaller time constants are actually better.
It can be shown that the feedback loop can operate even 
at $\tau \ll \Omega^{-1} \ll \Gamma^{-1}$; 
however, we are not interested in this regime because it requires 
a wide bandwidth of the control circuitry. Limiting ourselves to 
$\tau \sim S_I/(\Delta I)^2$, we see that the maximum achievable 
synchronization degree $D_{max}$ is about 90\% (that corresponds to the 
fidelity ${\cal F}$ of about 95\%). It is impossible 
to reach 100\% because the monitored simple phase estimate $\phi_m$
is significantly different from the actual $\phi$. It is interesting 
to note that a very crude estimate of $D_{max}$ as $\cos (\Delta \phi_{rms})$
using $\min (\Delta \phi_{rms}) \simeq 0.44$ from the analysis without 
feedback, works quite well, $\cos (0.44) =0.90$ (though for different $\tau$).
  Dashed line in Fig.\ \ref{D-F} shows the feedback performance for a 
classical signal corresponding to the dotted line in Fig.\ \ref{dphi-tau},
assuming $\tau (\Delta I)^2/S_I =1$. As expected, it operates much worse
than the quantum feedback because of the reason discussed above. 
[The crude estimate $D_{max} \sim \cos (\Delta \phi_{rms,\min}) =\cos (1.0)=
0.54$ still works well.]

        An important question is how the operation of the quantum feedback 
loop can be verified experimentally. One of the easiest ways is 
to check that the average value $\langle X\rangle$ of the in-phase 
quadrature component $X(t)$ becomes 
positive, while in absence of feedback ($F=0$) positive and negative
values of $X$ are obviously equally probable. Notice that {\it any}
Hamiltonian control of a qubit which is not based on the information
obtained from the detector (i.e.\ feedback control) cannot provide
nonzero $\langle X\rangle$. \cite{Xaver} 
It is easy to show that
$ \langle X\rangle = [D + \langle P \cos (2\Omega_m +\phi)\rangle ] 
        \tau \Delta I/4 $,
and since the second term in brackets vanishes at weak 
coupling (and $\varepsilon =0$), therefore 
$\langle X \rangle$ is directly related to $D$. 
The numerical results for $\langle X \rangle /(\tau \Delta I/4)$ 
practically coincide with the curves for $D$ in Fig.\ \ref{D-F} (within 
the thickness of the line).

\begin{figure}[t] 
\centerline{
\epsfxsize=2.7in 
\vspace{0.1cm}
\epsfbox{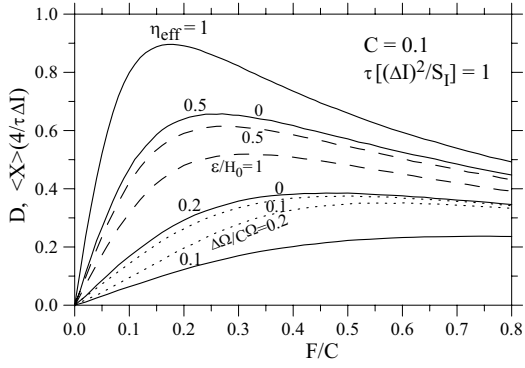} 
} 
\caption{Solid lines: synchronization degree $D$ (and in-phase current 
quadrature $\langle X\rangle$) as functions of $F$ for several 
values of the detection efficiency $\eta_{eff}$. Dashed and dotted lines
illustrate the effects of the energy mismatch ($\varepsilon \neq 0$) 
and the frequency mismatch ($\Omega \neq \Omega_0$). 
 }
\label{D-F-eta}\end{figure}

   The ideal case $\gamma =0$ is obviously not realizable in an 
experiment because of finite nonideality of a detector ($\eta <1$) 
and presence of an extra environment ($\gamma_e >0$). 
Both effects can be taken into account simultaneously 
introducing effective efficiency of quantum detection 
$\eta_{eff}=[\eta^{-1}+\gamma_e S_I/(\Delta I)^2]^{-1}$. 
Figure \ref{D-F-eta} shows (solid lines) the feedback performance 
for several values of $\eta_{eff}$ assuming $\tau (\Delta I)^2/S_I=1$.
One can see that $\eta_{eff} \sim 0.1$ is still a sufficient value 
for a noticeable operation of the quantum feedback loop. 
Notice that $D_{max}$ is limited at least by the state purity, $D_{max} < P$, 
which is \cite{witheta} $P \approx\sqrt{2\eta_{eff}}$ at $\eta_{eff}\ll 1$ 
and ${\cal C}/\eta \ll 1$ ($D_{max}=P$ can be reached by the feedback of 
Ref.\ \cite{Ruskov-fb} but not by the feedback studied here).

        Finally, let us discuss how accurately the conditions
$\Omega =\Omega_0$ and $\varepsilon =0$ should be satisfied in an experiment.
If $\Omega$ is different from $\Omega_0$, then without feedback 
the phase $\phi$ linearly grows in time [Eq.\ (\ref{dot-phi})]. 
However, if the feedback loop operation is faster than 
$|\Delta \Omega |=|\Omega-\Omega_0|$, 
the linear growth of $\phi$ is stopped by adjusting the Rabi frequency 
$\Omega_0$ to match the desired frequency $\Omega$. Dotted lines in Fig.\
\ref{D-F-eta} show the feedback operation for $\eta_{eff}=0.2$ and two
values of $\Delta \Omega$, confirming still good operation at 
$|\Delta \Omega| \ll {\cal C}\Omega \sim\Gamma \sim \tau^{-1}$.  
Notice that the frequency mismatch leads to nonzero $\langle \phi_m\rangle$
and therefore can be noticed and corrected. 
Energy mismatch ($\varepsilon \neq 0$) also worsens the performance
of the feedback loop; however, the dashed lines in Fig.\ \ref{D-F-eta} 
($\eta_{eff}=0.5$) show that a relatively large mismatch 
($\varepsilon \lesssim H_0$) can be tolerated.

    In conclusion, we have proposed and analyzed the quantum feedback
loop for a solid-state qubit, based on monitoring the phase
of Rabi oscillations via quadrature components of the current in a
weakly coupled detector. Surprisingly, it operates much better than 
one could guess from relatively small spectral peak of the detector
output in absence of feedback.
The work was supported by NSA and ARDA under ARO grant DAAD19-01-1-0491.



\begin{references}
\bibitem{Rabi} We use the term ``Rabi oscillations'' for 
        evolution of a qubit which state is not an energy eigenstate, 
        not implying an rf field presence.
\bibitem{witheta} At finite $\eta$ (assuming $\varepsilon =0$ and 
$\Omega =\Omega_0$) the phase equation is 
$\dot{\phi}= -[I(t)-I_0] \sin (\Omega t 
+\phi )(\Delta I/S_I)/P^2  -(\gamma /2) \sin (2\Omega t +2\phi)$. At  
${\cal C}/\eta \ll 1$ the last term can be neglected and state purity 
$P\equiv [(\rho_{11}-\rho_{22})^2+(2\mbox{Im}\rho_{12})^2]^{1/2}$ 
is given by $P^2 = 1+1/2\eta -\sqrt{(1+1/2\eta )^2-2}$. In particular,
$P\approx \sqrt{2\eta}$ at $\eta \ll 1$.
\bibitem{H/Hfb} Notice that $H_{fb}/H_0 \ll 1$ because $F/{\cal C}\lesssim 1$
  and ${\cal C}\ll 1$, so the feedback requires only a weak change of 
    the tunnel barrier between the qubit states.
\bibitem{Xaver} Some procedures which include periodic dissipation phase 
        (nonunitary operation) can be used to achieve nonzero 
        $\langle X\rangle$, though much less efficiently than by the 
        quantum feedback.
\end{references}
\end{document}